\documentclass[twoside]{article}
\usepackage{fleqn,espcrc2modif}

\usepackage{graphicx}
\usepackage[figuresright]{rotating}

\newcommand{\AmS}{{\protect\the\textfont2
  A\kern-.1667em\lower.5ex\hbox{M}\kern-.125emS}}

\def\be{\begin{equation}}
\def\ee{\end{equation}}
\def\bea{\begin{eqnarray}}
\def\eea{\end{eqnarray}}
\def\ba{\begin{array}} %%%%%%%
\def\ea{\end{array}}   %%%%%%%

\def\simge{\mathrel{%
   \rlap{\raise 0.511ex \hbox{$>$}}{\lower 0.511ex \hbox{$\sim$}}}}
\def\simle{\mathrel{
   \rlap{\raise 0.511ex \hbox{$<$}}{\lower 0.511ex \hbox{$\sim$}}}}

\title{\LARGE \bf ABOUT \,THE \,ORIGINS 
\,OF \,THE   \medskip\smallskip \\
SUPERSYMMETRIC \,STANDARD \,MODEL\bigskip\bigskip \\}

\author{\Large P. Fayet\,\vspace{.8pc}\address[LPTENS]{
Laboratoire de Physique Th\'eorique de l'Ecole Normale 
Sup\'erieure \\
\ \ \,24 rue Lhomond, 75231 Paris Cedex 05, France
}
\thanks{UMR 8549, Unit\'e Mixte du CNRS et de l'Ecole Normale 
Sup\'erieure. \hskip 5.5truecm\ \hbox{\ }
\hbox{\ \ \ \ \ } \hskip 2truecm {\normalsize LPTENS/01-34}}
}
       
\begin{document}

\begin{abstract}

\vspace{1pc}

\begin{center}
\large
Invited talk at the International Symposium \ \ ``30 Years of Supersymmetry'' 
\\
\vskip .1truecm
Minneapolis, \ Oct. 13-15, \,2000
\end{center}

\vskip .2truecm

Could one use supersymmetry to relate the fermions,
constituants of matter, with the bosons messengers of the interactions\,?
This is, ideally, what a symmetry between fermions and bosons would 
be expected to do.

\vskip .1truecm

However many obstacles seemed, long ago, to prevent
supersymmetry from possibly being a fundamental symmetry of Nature.
Which fermions and bosons could be related\,?
Is spontaneous supersymmetry breaking possible at all\,?
If yes, where is the corresponding spin-$\frac{1}{2}\,$ Goldstone fermion\,?
Supersymmetric theories also involve Majorana fermions, unknown in Nature. 
And how could we define conserved quantum numbers like $B$ and $\,L$, 
\,when these are carried by fundamental (Dirac) fermions only, not by bosons\,?

\vskip .1truecm
An early attempt to relate the photon with a ``neutrino''
led us to $\,R$-invariance and to a new \hbox{$R\,$ quantum} number 
carried by the supersymmetry generator,
but this ``neutrino'' had to be reinterpreted as a new particle, 
the {\it \,photino}. 
We also had to introduce bosons carrying ``fermion numbers'' 
$\,B\,$ and $\,L$, 
\,which became the squarks and sleptons.
This led to the Supersymmetric Standard Model, 
involving $\,SU(3)\times SU(2)\times U(1)\,$ gauge superfields
interacting with chiral quark and lepton superfields, 
and {\it \,two\,} doublet Higgs superfields responsible for 
quark and lepton masses.
$\,R$-parity, deeply related with $\,B\,$ and $\,L\,$ 
conservation laws, appeared as a remnant 
of the original $R$-invariance, 
reduced to a discrete symmetry so that the gravitino and gluinos 
can acquire masses. 
We also comment about supersymmetry breaking.

\vspace{1pc}
\end{abstract}

\maketitle

\section{GENERAL OVERVIEW}
\label{sec:intro}

\vskip .1truecm

Where is the idea of a ``Superworld'' coming from\,? 
Could half of the particles, at least, 
have escaped our direct observations\,?

\vskip .25truecm

According to common wisdom supersymmetry is an algebraic structure
which allows in principle to relate 
particles of half-integer spins, namely fermions, 
with particles of integer spins, which are bosons.
Could one relate the fermions, {\it \,constituants of matter}, 
with the bosons, which appear 
as the {\it \,messengers of interactions}, 
and then arrive at some sort of unification between Matter and Forces\,?
This would certainly be very attractive, but unfortunately things 
are not so simple.

\vskip .25truecm

Let's go back in time, about thirty years ago.
The algebraic structure of supersymmetry in four dimensions,
whose origins in the East and in the West were related in details 
at this Conference,
was introduced in the beginning of the seventies by 
Gol'fand and Likhtman~\cite{gl},
Volkov and Akulov~\cite{va}, and Wess and Zumino~\cite{wz}.
It involves a spin-$\frac{1}{2}\,$ fermionic symmetry generator
relating fermionic with bosonic fields, 
in relativistic quantum field theories.

\vskip .35truecm

Supersymmetry essentially appeared, at the beginning, 
as a rather formal mathematical structure.
At that time it was not at all clear if, and even less how,
supersymmetry could have a chance to actually relate fermions with bosons, 
in a physical theory of particles. Even more, although supersymmetry 
is commonly known as relating fermions with bosons, its algebra 
does not even require the existence of elementary bosons at all\,! 
\,(In the framework of non-linear supersymmetry a fermionic field 
can be turned 
into a composite bosonic field made of two fermionic ones~\hbox{\cite{va}}.)
\,In any case, supersymmetry as an algebraic structure does not by itself 
require the existence of superpartners for all particles,
to which we are so accustomed now.

\vskip .35truecm
Considering an algebra is not sufficient to give it a physical meaning. 
The existence of $\,SU(2)\,$ 
as a mathematical structure does not mean isospin or weak-interaction 
symmetries, charm is not a consequence of $\,SU(4),\,$
nor grand-unification a prediction of $\,SU(5)$ ...
An algebra may be of interest, in physics,
depending on if and how it can actually be used to describe
the real world. In the early days of supersymmetry the hopes 
were more like using it as a tool to understand better the properties of 
relativistic field theories; or trying to relate amplitudes 
involving integer-spin mesons with those involving half-integer-spin baryons; 
or at best, if supersymmetry could actually be used in the description 
of the real world and were to be realized at a fundamental level, 
to attempt to use it to relate spin-1 gluons with spin-$\frac{1}{2}$ quarks, 
for example.
However, while very interesting 
from the point of view of relativistic field theory, 
supersymmetry seemed, in the early days, clearly inappropriate 
for a description of our physical world, for obvious and less obvious reasons, 
which often tend to be somewhat forgotten, 
now that we got so accustomed to deal with Supersymmetric 
extensions of the Standard Model.

\vskip .25truecm

At first, one could not identify physical bosons and fermions 
that might be related under such a symmetry, in contrast
with other symmetries which relate known particles together, 
such as isospin, electroweak-interaction or grand-unification 
symmetries. Even if it still remains hypothetical, the grand-unification 
symmetry, for example, should relate known leptons with known quarks, 
while supersymmetry does not seem to allow 
for similar relations between known bosons and known fermions.

\vskip .25truecm
It even seemed initially
that {\it supersymmetry could not be spontaneously broken at all\,} 
\,-- also in contrast with ordinary symmetries like internal 
or gauge symmetries --\,
which would imply that bosons and fermions be systematically degenerated 
in mass\,! Unless of course supersymmetry-breaking terms are explicitly 
introduced ``by hand'', which would spoil the fundamental r\^ole of the 
supersymmetry and prevent it from being realized as a local gauge symmetry.
\,In any case, we know significantly more fundamental 
fermion fields, describing leptons and quarks, than boson fields. 
To help better set the stage, let us recall that at the time, in 1974, 
only two fermion families were known and not even complete 
with the charm quark still to be discovered, 
neutral current effects had just been discovered the year before, in 1973,
with very little information available
about the structure of the weak neutral current(s?), and the 
lower limit on the mass of a charged $\,W\,$ boson
was something like \,5 \,GeV$/c^2$. 
The Standard Model was a recent theoretical construction, far from ``Standard''
in today's sense;
its $\,W^\pm$ and $Z$ bosons, of course hypothetical, were 
considered as really very heavy; 
\,and even-more-hypothetical Higgs fields were generally viewed 
as a technical device to trigger or mimic the 
spontaneous breaking of the gauge symmetry.

\vskip .25truecm
Furthermore, independently of the previous problems of supersymmetry, 
these theories also involve, systematically, {\it \,self-conjugate 
Majorana spinors,} \,while Majorana fermions are completely unknown in Nature
%\hbox{\,-- completely unknown in Nature --\,} 
%while 
\,-- the fermions that we know 
all appearing 
as {\it Dirac fermions\,} carrying {\it \,conserved quantum numbers}, 
$\,B\,$ and $\,L$.
\,Speaking of $\,B\,$ and $\,L$,
\,how could we account for the conservation of these 
``fermionic numbers'' 
%$\,B\,$ and $\,L\,$ 
(only carried by fermions) \,in a supersymmetric theory,
in which fermions are related to bosons\,?
Should some bosons carry fermionic number 
\,-- and what about ``fermion number'' conservation, especially if such bosons 
carrying ``fermion number'' could be exchanged between quarks and leptons\,?
In view of all these problems, or even without being explicitly aware of them, 
most physicists considered supersymmetry 
as irrelevant for ``real physics''.

\vskip .25truecm
Still this algebraic structure could be taken seriously 
as a possible symmetry of the physics 
of fundamental particles and interactions, once we understood 
that the above obstacles preventing the application of supersymmetry 
to the real world could be overcome. 
In particular the last questions about the conservation of quantum numbers 
led us to consider the possibilities of having an additive 
quantum number carried by the supersymmetry generator 
\ -- and this was the $\,R$ quantum number 
associated with the continuous \,``$R$-invariance'',
also implying restrictions on the allowed superpotential \,--\,
and to have bosons carrying ``fermion numbers'' $\,B\,$ and $\,L$, 
\,which became the squarks and sleptons.

\vskip .25truecm
After an initial attempt illustrating how far one could go in trying
to relate known particles together
(in particular the photon with a ``neutrino'', 
and the $W^\pm$'s with charged ``leptons'',  \linebreak
also related with charged Higgs bosons $H^\pm$),
\,and the limitations of this approach,
in a $\,SU(2) \times U(1)\,$ 
electroweak theory involving 
two doublet Higgs superfields now known as 
$\,H_1\,$ and $\,H_2\,$~\cite{R}, 
~we were quickly led to reinterpret the fermions of this model 
\,(which all possess $\,\pm 1\,$ unit of a conserved additive 
$\,R\,$ quantum number carried by the supersymmetry generator)
\,as belonging to a new class of particles.
The ``neutrino'' ought to be considered as a really new particle,
a ``photonic neutrino'', \,a name I transformed in 1977 
into {\it \,photino}; the fermionic partners of the colored gluons 
(quite distinct from the quarks) then becoming the {\it \,gluinos}, 
\,and so on.
More generally this led us to postulate the existence of new 
$\,R$-odd ``superpartners'' for all particles and consider them seriously, 
despite their rather non-conventional properties:
e.g. new bosons carrying ``fermion'' number, now known 
as {\it \,sleptons} and {\it \,squarks}, \,or Majorana fermions 
transforming as an $\,SU(3)\,$ color octet, 
which are precisely the {\it \,gluinos}, etc..
In addition the electroweak breaking must be induced by \hbox{\it a pair\,} 
of electroweak Higgs doublets, not just a single one as 
in the Standard Model, 
which requires the existence of {\it \,charged Higgs bosons},
\,and of several neutral ones~\cite{ssm,grav}.

\vskip .25truecm

This construction illustrates that supersymmetry is the framework 
in which fundamental \hbox{spin-0} (Englert-Brout) Higgs fields 
find their natural place.
$\!$In ordinary gauge theories spin-0 Higgs fields 
were generally considered as ad hoc additions to the sectors 
of spin-1 gauge bosons and spin-$\,\frac{1}{2}$ fermions, 
and much effort was subsequently devoted at attempting to avoid them, 
as in technicolor theories for example 
(which still could not get rid of them completely,
independently of other problems).
In this framework of supersymmetry fundamental spin-0 fields 
are taken seriously from the beginning, 
exactly on the same footing as spin-$\frac{1}{2}$ 
fields which appear as their fermionic counterparts under supersymmetry.
The number of {\it \,categories of fundamental objects\,}
is decreased from \,3\, in 
ordinary gauge theories (spin-1 gauge bosons interacting with
spin-$\frac{1}{2}\,$ fermions and spin-0 Higgs fields) 
\,to \,2\, only in supersymmetric gauge theories: 
\,(\hbox{spin-1}/ {spin-$\frac{1}{2}$})\, multiplets of ``gauge particles'' 
interacting with \,(spin-$\frac{1}{2}$/spin-0)\, ``chiral'' multiplets 
describing spin-0 Higgs fields as well as spin-$\frac{1}{2}\,$ lepton and quark 
fields \,-- although they are not directly related by the supersymmetry.
(Indeed a new discrete symmetry known as $\,R$-parity comes in to distinguish,
among \,\,spin-$\frac{1}{2}$/spin-0\,\, chiral multiplets, 
the two separate sectors of Higgs multiplets 
on one hand, and quark and lepton multiplets, on the other hand.)

\vskip .25truecm
In the same manner the number of {\it \,categories\, of 
coupling constants\,}
gets also reduced down to \,2, namely gauge and Yukawa couplings only.
The Higgs potential in a supersymmetric theory is now determined 
by the gauge and Yukawa couplings of spin-$\frac{1}{2}\,$ particles,
with Higgs mass parameters identical to fermion mass parameters 
up to supersymmetry-breaking contributions (naturally expected to be
$\,\simle\,$ electroweak scale \,\,given their interplay with Higgs v.e.v.'s,
if no excessive fine-tuning is to be performed).
In particular, the quartic couplings of the two Higgs doublets 
described by the superfields $\,H_1$ and $\,H_2\,$ mentioned earlier, 
as they appear for example in the ``minimal'' version of the Supersymmetric 
Standard Model, are completely fixed by the supersymmetry
in terms of $\,g^2$ and $\,g^2+g'^2$ \,($\,g\,$ and $\,g'\,$
being the $\,SU(2)\times\,U(1)$ electroweak gauge couplings),
a fact at the origin of various mass relations 
involving massive spin-1 gauge bosons and spin-0 Higgs bosons.

\vskip .25truecm

The still-hypothetical superpartners may be 
distinguished by a new quantum number called $R\,$-parity, 
\,first defined in terms of the previous $R\,$ quantum number
as $\,R_p=(-1)^R$, ~i.e. $\,+1$ 
for the ordinary particles and $\,-1\,$ for their superpartners. 
It is associated with a $\,Z_2$ remnant of the previous $\,R$-symmetry
acting continuously on gauge, lepton, quark and Higgs 
superfields as in \,\cite{ssm},
\,which must be be abandoned as a continuous symmetry so as to allow for the 
gravitino~\cite{grav} and gluinos~\cite{rp} \,-- upon which 
the continuous $\,R$-symmetry acts chirally --\,
\,to acquire masses 
~(whatever is the actual mechanism, still unknown,  
ultimately responsible for the spontaneous 
breaking of the supersymmetry).
This new discrete quantum number defined as $\,R$-parity~\hbox{\cite{rp}}
may be multiplicatively conserved in a natural way, and is especially useful 
to guarantee the absence of 
unwanted interactions mediated by squark or slepton exchanges \,--\, 
in connection with the fact that $\,R$-symmetries imply restrictions 
on the allowed superpotential, actually permitting, in the present case, 
all useful superpotential terms necessary to generate quark and lepton masses, 
while excluding dangerous $\,B\,$ and $\,L\,$-violating terms.
The conservation \,(or non-conservation)\,
of $\,R$-parity is therefore closely related with the conservation 
\,(or non-conservation)\, of baryon and lepton numbers, 
$\,B\,$ and $\,L$, ~as illustrated by the well-known formula 
reexpressing $\,R$-parity in terms of baryon and lepton numbers, 
as $\,(-1)\,^{2S} \ (-1)\,^{3B+L}$
~\cite{ff}.

\vskip .3truecm
The finding of the basic building blocks of 
the Supersymmetric Standard Model,
whe\-ther ``minimal'' or not, allowed for the 
experimental searches for ``supersymmetric particles'', 
which started with the first searches for gluinos and photinos, 
selectrons and smuons, in the years 1978-1980,
and have been going on continuously since.
These searches often use the ``missing energy'' signature
corresponding to energy-momentum carried away by unobserved 
neutralinos~\cite{ssm,ff,ff2,ff3}.
A conserved $R$-parity also ensures the stability 
of the ``lightest supersymmetric particle'',
\,a good candidate to constitute the non-baryonic Dark Matter 
that seems to be present in the Universe.

\vskip .25truecm
The general opinion of the scientific community 
towards supersymmetry and supersymmetric extensions of the Standard Model 
has considerably changed since the early days, 
in view of all the nice features of such theories, 
including their fundamental relation with gravity, 
%their link with extra spacetime dimensions and 
their improved renormalisation properties (with the softening or
cancellation of divergencies between boson and fermion contributions),
the effects of the new particles on the high-energy evolution 
of gauge couplings, 
the r\^ole of supersymmetry in the consistency of (super)string theories, etc..
And it is now widely admitted that supersymmetry may 
well be the next fundamental symmetry to be discovered in the physics of 
fundamental particles and interactions, 
although this remains to be experimentally proven.

\vspace{3mm}

\section{NATURE DOES NOT SEEM TO BE SUPERSYMMETRIC\,!}
\label{sec:na}

\vskip .05truecm

The supersymmetry algebra 
\bea
\label{alg}
\cases{ \ \ \ba{cccc}
\{ \ \,Q , \ {\bar Q} \ \,\} &=& 
- \ 2\ \,\gamma_{\mu} \, P^{\mu} &, \vspace {0.3 true cm} \cr 
[ \,\ Q, \ P^{\mu} \ ] &=& \ \ \ \ \ \ 0  \ \ \ &,
\ea
}\label{ss}           
\eea
was introduced, 
in the years 1971-1973, by three different groups, 
with quite different motivations. 
Gol'fand and Likhtman~\cite{gl} first introduced it 
with the hope of understanding parity-violation:
when the Majorana supersymmetry generator
is written as a two-component 
chiral Dirac spinor (say $\,Q_L$), \,one may have the impression that the 
supersymmetry algebra, which then involves a chiral projector 
in the right-handside of the anticommutation relation (\ref{alg}), 
is intrinsically parity-violating; 
%\,(which, however, is not the case); 
they suggested that such models must therefore necessarily violate parity, 
probably thinking this could lead to an explanation 
for parity-violation in weak interactions.
Volkov and Akulov~\cite{va} hoped to explain the masslessness of the neutrino 
from a possible interpretation as a spin-$\frac{1}{2}\,$
Goldstone particle, while Wess 
and Zumino~\cite{wz} wrote the algebra by extending to four dimensions the 
``supergauge'' (i.e. supersymmetry) transformations~\cite{2d}, 
and algebra~\hbox{\cite{2dalg}},
acting on the two-dimensional string worldsheet.
However, the mathematical existence of an algebraic structure does 
not imply that it has to play a r\^ole as an invariance 
of the fundamental laws of Nature.

\vskip .3truecm

Indeed many obstacles seemed, long ago, to prevent supersymmetry 
from possibly being a fundamental symmetry of Nature.
Which bosons and fermions could be related by supersymmetry\,?
May be supersymmetry could act at the level of composite objects, e.g. 
$\!\!$as relating \linebreak baryons with mesons\,?
Or should it act at a fundamental level, i.e. 
$\!$at the level of quarks and gluons\,? But quarks are color triplets, 
and electrically charged,
while gluons transform as an $\,SU(3)\,$ color octet, 
and are electrically neutral\,! 
\,Is spontaneous supersymmetry breaking possible at all\,?
\,If yes, 
where is the spin-$\frac{1}{2}\,$ Goldstone fermion of supersymmetry, 
if not one of the known neutrinos\,?
Can we use supersymmetry to relate directly known bosons and fermions\,?
And, if not, why\,?
\,If known bosons and fermions cannot be directly related by supersymmetry, 
do we have to accept this as the sign that supersymmetry is {\it \,not\,}
a symmetry of the fundamental laws of Nature\,?
If we still insist to work within the framework of supersymmetry,
how could it be possible to define conserved baryon and lepton numbers 
in such theories,
which systematically involve {\it \,self-conjugate\,}
Majorana fermions, unknown in Nature, while $\,B\,$ and $\,L\,$ 
are carried only by fundamental (Dirac) fermions \,-- not by bosons~? 
And, once we are finally led to postulate the existence of new bosons 
carrying $\,B\,$ and $\,L\,$ 
\,-- the new spin-0 squarks and sleptons --\,
can we prevent them from mediating new unwanted interactions,
which would have disastrous effects\,? 

\vskip .3truecm

While bosons and fermions should have equal masses 
in a supersymmetric theory, this is certainly not the case in Nature.
Supersymmetry should then clearly be broken.
But spontaneous supersymmetry breaking is notoriously difficult to achieve, 
to the point that it was even initially thought to be impossible\,!
Why is it so\,?  Here it is important to point out an important difference 
between supersymmetry and other symmetries such as internal 
or gauge symmetries for example. To break spontaneously 
such ``ordinary'' symmetries, 
it is sufficient to arrange so that the symmetry-preserving would-be 
vacuum state has more energy than symmetry-breaking vacua,
so that it gets unstable. This is for example how the electroweak symmetry 
gets spontaneously broken in the Standard Model,
as soon as the Higgs mass parameter $\,\mu^2\,$ is taken to be negative.

\vskip .3truecm

But in supersymmetry one no longer has the same freedom 
to fix at will the potential of scalar fields as in ordinary gauge theories,
since this potential is now largely determined by gauge and Yukawa couplings.
Furthermore supersymmetry is in fact a very special symmetry, 
since the Hamiltonian, which governs the determination of the
vacuum state through the minimization of the potential, 
is also directly related 
by the supersymmetry algebra itself to the supersymmetry
generator, precisely the one that we would like to see spontaneously broken.
Actually this hamiltonian $\,H$, \,which appears in the right-handside 
of the anticommutation relations  (\ref{alg}), can be expressed 
proportionally to the sum of the squares of the components 
of the supersymmetry generator, as
$\,H=\frac{1}{4}\ \sum_\alpha\,Q_\alpha^{\ 2}\,$.
\,This implies that a supersymmetry preserving vacuum state must 
have vanishing energy, while any candidate for a ``vacuum state'' 
which would not be invariant under supersymmetry
may na\"{\i}vely be expected 
to have a larger, positive, energy~\cite{iz}~\footnote{Such a would-be 
supersymmetry breaking state corresponds, in global supersymmetry,
to a {\it \,strictly positive\,} energy density 
\,-- the scalar potential being expressed proportionally 
to the sum of the squares of the auxiliary 
$\,D, \ F\,$ and $\,G\,$ components, as
$\ \,V\,=\,\frac{1}{2}\ \sum\ (\,D^2\,+\,F^2\,+\,G^2\,)\ \,$.}. 
As a result, potential candidates for
supersymmetry breaking vacuum states
seemed to be necessarily unstable, leading to the question:
\be
\ \ \ba{c}
\hbox {\it{Is spontaneous supersymmetry breaking}} \vspace{.5mm}\\ 
\hbox {\it{possible at all\ ?}}
\ea
\ee
In spite of the above argument,
several ways of breaking spontaneously 
global or local supersymmetry have been found.
%~\cite{fi,F,crem}.\,
But spontaneous supersymmetry breaking remains in general
rather difficult to obtain, at least for global supersymmetry, 
due to the strong tendency of such theories 
to resist spontaneous supersymmetry breaking
by preferring systematically supersymmetric vacua. 
Only in very exceptional situations can the existence of 
such vacua be completely avoided\,! Gauge symmetries, on the other hand, 
get rather easily (even sometimes too easily) \,spontaneously broken, 
in supersymmetric theories.

\vskip .3truecm
In global supersymmetry a non-supersymme\-tric state has, 
in principle, 
always more energy than a supersymmetric one; 
it then seems that it should always be unstable\,!
%the stable vacuum state being necessarily supersymmetric\,!
\,Still it is possible to escape this general result 
\,-- and this is the key to spontaneous supersymmetry breaking --\,
if one can arrange to be in one of those rare situations 
for which {\it \,no supersymmetric 
state\,} {\it exists at all\,}
\,-- the set of equations for the auxiliary field v.e.v.'s 
$\ <\!D\!>'\hbox{s}\,=\,\,<\!F\!>'\hbox{s}\,=\,\hbox{$<\!G\!>'\hbox{s}$}\,=\,0\ $ 
having {\it \,no solution at all\,}.
\,But these situations are in general 
quite exceptional. (This is in sharp contrast 
with ordinary symmetries, in particular gauge symmetries,
for which it is sufficient to arrange for non-symmetric states 
to have less energy than symmetric ones, which is easy to achieve.)
These rare situations usually involve an abelian $\,U(1)$ 
gauge group~\cite{fi}, 
allowing for a linear \,``$\,\xi\,D\,$''\, 
term in the Lagrangian density\,\footnote{Even 
in the presence of such a term, one frequently still does not get a
spontaneous breaking of the supersymmetry: one has to be very careful 
so as to avoid the presence
of supersymmetry restoring vacuum states with vanishing energy, 
which generally tend to exist.
};
\,and/or an appropriate set of chiral superfields 
with special superpotential interactions
which must be very carefully chosen~\cite{F}, 
preferentially with the help of additional symmetries such as $\,R$-symmetries.
\,In local supersymmetry~\cite{sugra}, which includes gravity,
one also has to arrange, at the price of a very severe fine-tuning,
for the energy density of the vacuum to vanish exactly~\cite{crem}, 
or almost exactly, to an extremely good accuracy,
so as not to generate an unacceptably large value of the 
cosmological constant $\,\Lambda\,$.

\vskip .5truecm

Whatever the mechanism of supersymmetry breaking, we still have to get
a physical world which looks like ours (which will lead 
to postulate the existence of the superpartners). 
%for all ordinary particles).
\,Of course just accepting the possibility of explicit supersymmetry breaking 
without worrying too much about the origin of 
supersymmetry breaking terms,
as is frequently done now, makes things much easier
\,-- but also at the price of introducing a large number 
of arbitrary parameters, 
coefficients of these supersymmetry breaking terms.
In any case such terms must have their origin 
in a spontaneous supersymmetry breaking mechanism, 
if we want supersymmetry to play a fundamental r\^ole, 
especially if it is to be realized as a local fermionic gauge symmetry, 
as in supergravity theories. 
We shall come back to this question of supersymmetry breaking later. 
In between, we note that the spontaneous breaking of the global supersymmetry 
must in any case generate a massless spin-$\frac{1}{2}\,$ Goldstone particle, 
leading to the next question,
\be
\ \ \ba{c}
\hbox {\it{Where is the spin-$\frac{1}{2}\,$ 
Goldstone fermion}} \vspace{.5mm} \\
\hbox {\it{of supersymmetry\ ?}} \vspace{.2mm} \\
\ea
\ee

\noindent
Could it be one of the known neutrinos~\cite{va}\,?
A first attempt at implementing this idea 
within a $\,SU(2) \times U(1)\,$ electroweak model of 
``leptons''~\hbox{\cite{R}}
quickly illustrated that it could not be pursued very far. 
The ``leptons'' of this first electroweak model 
were soon reinterpreted to become the ``charginos'' and ``neutralinos'' 
of the Supersymmetric Standard Model.

\vskip .32truecm

If the Goldstone fermion of supersymmetry is not one of the known
neutrinos, why hasn't it been observed\,?
Today we tend not to think at all about the question,
since: 1) the generalized use of soft terms breaking {\it \,explicitly\,}
the supersymmetry seems to make this question irrelevant; \ 
2) 
since supersymmetry has to be realized locally anyway, 
within the framework of supergravity~\cite{sugra}, 
the massless \hbox{spin-$\frac{1}{2}\,$} Goldstone fermion (``goldstino'') 
should in any case be eliminated 
in favor of extra degrees of freedom for a massive 
spin-$\frac{3}{2}\,$ gravitino~\cite{grav,crem}.

\vskip .33truecm
But where is the gravitino, and why has no one ever seen 
a fundamental spin-$\frac{3}{2}$ particle\,?
Should this already be taken as an argument against supersymmetry 
and supergravity theories\,? Indeed should one consider that the crucial test 
of such theories should be the discovery of a new fundamental 
spin-$\frac{3}{2}\,$ particle\,?
In that case, how could it manifest its presence\,?

\vskip .3truecm
To discuss this question properly we need to know how this 
spin-$\frac{3}{2}$ particle should couple to the other ones, 
which requires us to know how bosons and fermions 
could be associated under supersymmetry~\cite{ssm}.
In any case, we might already anticipate 
that the interactions of the gravitino, 
with amplitudes proportional to the square root of the Newton constant  
$\,\sqrt{G_N} \simeq 10^{-19}\ \,\hbox{GeV}^{-1}$, 
~should be absolutely negligible in particle physics.
Quite surprisingly this may, however, not necessarily be true,
despite the extreme smallness of Newton's constant\,!
If the spin-$\frac{3}{2}$ gravitino turns out to be light 
(which is the case if supersymmetry is broken ``at the electroweak scale'' 
or even at some larger intermediate scale) 
\,it would still interact very much like the massless spin-$\frac{1}{2}$ 
goldstino of global supersymmetry, 
according to the ``equi\-valence theorem'' of supersymmetry~\cite{grav}.
We are then back again to our initial question, 
where is the \hbox{spin-$\frac{1}{2}$}\, Goldstone fermion of supersymmetry\,?
But we are now in a position to answer, 
the direct detectability of the gravitino depending crucially on the value 
of its mass $\,m_{3/2}\,$, \,itself fixed 
\,(according to the relation $\,m_{3/2}=\kappa d/\sqrt6\,$) \linebreak
\,by the supersymmetry breaking scale 
$\sqrt d = \Lambda_{ss} \approx
\sqrt{\phantom d \!\! m_{3/2}\,m_{\hbox{\footnotesize Planck}}}
\ $~\cite{grav,grav2}.
In particular the gravitino gets essentially ``invisible'' 
in particle physics experiments, as soon as the supersymmetry breaking scale 
is large enough, compared to the electroweak scale.
This seems indeed the most likely situation.
On the other hand, a sufficiently light gravitino could be directly 
detectable in particle physics experiments\,!

\vskip .5truecm

In any case, irrespective of the question of supersymmetry breaking, 
the crucial question,
if supersymmetry is to be relevant in particle physics, is:
\be
\ \ \ \ba{c}
\hbox {\it{Which bosons and fermions}}  \vspace{.5mm} \\
\hbox {\it{could be related by supersymmetry\,?}}
\ea
\ee
But there seems to be no answer since known bosons
and fermions do not appear to have much in common 
\,-- excepted, maybe, for the photon and the neutrino. 
This track deserved to be explored~\cite{R},  
but one cannot really go very far in this direction.
In a more general way the number of (known) degrees of freedom 
is significantly larger for the fermions 
(now 90, for three families of quarks and leptons)
than for the bosons (27 \,for the gluons, the photon 
and the $\,W^\pm$ and $\,Z\,$ bosons, ignoring for the moment
the spin-2 graviton and the not-yet-discovered Higgs boson).
And these fermions and bosons have very different gauge symmetry 
properties\,!

\vskip .3truecm

As we have already mentioned,
supersymmetric theories also involve, systematically, 
self-conjugate Majorana spinors \,-- unobserved in Nature --\, 
while the fermions that we know all appear 
as Dirac fermions carrying conserved $\,B\,$ and $\,L\,$
quantum numbers.
This leads to the question

\be
\ \ \ \ \ \ba{c} 
\hbox {\it How could one define (conserved)}  \vspace{.5mm} \\
\hbox {\it baryon and lepton numbers,} \vspace{.5mm} \\
\hbox{\it in a supersymmetric theory\ ?}
\ea
\ee

These quantum numbers, presently known to be
carried by fundamental fermions only, not by bosons, 
seem to appear in Nature as {\it intrinsically-fermionic\,} numbers.
Such a feature cannot be maintained in a supersymmetric 
theory, and one has to accept the (then rather heretic)
idea of attributing 
baryon and lepton numbers to fundamental bosons, as well as to fermions.
These new bosons carrying $\,B\,$ or $\,L\,$ are
the superpartners of the spin-$\frac{1}{2}$ quarks and leptons, namely
the now-familiar (although still unobserved)
spin-0  {\it \,squarks\,} and {\it \,sleptons\,}.
Altogether, all known particles should be associated 
with new {\it \,superpartners\,}~\cite{ssm}.

\vskip .3truecm

Nowadays we are so used to deal with spin-0 squarks and sleptons, 
carrying baryon and lepton numbers almost by definition, 
that we can hardly imagine this could once have appeared as a problem.
Its solution went through accepting the idea 
of attributing baryon or lepton numbers to a large number of new 
fundamental bosons.
But if such new spin-0 squarks and sleptons are introduced,
their direct (Yukawa) exchanges between ordinary 
 quarks and leptons, if allowed, 
could lead to an immediate disaster, preventing us from getting a theory 
of electroweak and strong interactions mediated by spin-1 
gauge bosons only (and not spin-0 particles), 
with conserved $\,B\,$ and $\,L\,$ quantum numbers\,! 
\bea 
\ \ \ba{c} 
\hbox {\it{How can we avoid unwanted interactions}} \vspace{.5mm}\\ 
\hbox{\it{mediated by spin-0 squark and slepton}} \vspace{.5mm}\\
\hbox{\it{exchanges\,?}} \vspace{.2mm}
\ea
\eea

\noindent
Fortunately, we can naturally avoid 
such unwanted interactions, thanks to $\,R$-parity
(a discrete remnant of the continuous $\,U(1)\ R$-symmetry) which, if present,
guarantees that squarks and sleptons can{\it not\,} be
directly exchanged between ordinary quarks and leptons, 
allowing for conserved baryon and lepton numbers 
in supersymmetric theories.

\vspace{5mm}

\section{$R\,$-INVARIANCE, AND ELECTRO\-WEAK BREAKING}
\label{sec:R}

\vskip .05truecm

Let us now return to an early attempt at relating 
{\it \,existing\,} bosons and fermions together~\cite{R},
also at the origin of the definition 
of the continuous $\,R\,$-invariance\,\footnote{\label{footnote:R}This model is reminiscent of
a presupersymmetry model involving two Higgs doublets 
and (associated) fermion doublets, 
with Yukawa and $\,\varphi^4\,$ interactions already restricted 
by a continuous $\,Q$-invariance in a way which prepares for these
Higgs and fermion doublets to get related by a supersymmetry~\cite{P}. 
One unit of $\,Q\,$ is then carried by the supersymmetry generator.
These $\,Q\,$ transformations act on gauge and Higgs superfields as follows:
$$
\left\{ \ 
\ba{ccc}
V(\,x,\,\theta,\,\bar\theta\ ) &\ \,\stackrel{Q}{\longrightarrow} \,\ &  
\ \ V (\,x, \,\theta \,e^{-i\alpha},\,\bar\theta \, e^{i\alpha}\,) \ \ ,  
\\ [.0truecm]
H_{1,2} \,(\,x,\,\theta\,) &\stackrel{Q}{\longrightarrow}  & e^{i\alpha} \ \ 
H_{1,2}\, (\,x, \,\theta \,e^{-i\alpha}\,)  \ \ ,  
\ea  \right.
$$
allowing for a direct $\ \mu \ H_1 H_2\ $ Higgs superfield
mass term in the superpotential.
The definition of $\,Q\,$-invariance was then modified, 
so that it survives the spontaneous breaking 
of the electroweak symmetry~\cite{R}.
This led to $\,R\,$-invariance, 
acting as follows:
$$
\left\{ \ 
\ba{ccc}
V(\,x,\,\theta,\,\bar\theta\ ) &\ \,\stackrel{R}{\longrightarrow} \,\ &  
\ \ V (\,x, \,\theta \,e^{-i\alpha},\,\bar\theta \, e^{i\alpha}\,)\ \ ,  
\\ 
H_{1,2} \,(\,x,\,\theta\,) &\stackrel{R}{\longrightarrow}  &\ \ 
H_{1,2}\, (\,x, \,\theta \,e^{-i\alpha}\,)  \ \ .
\ea  \right.
$$
The direct $\,\mu\ H_1 H_2\,$ mass term
\,(with $\,R$-index $\,n=0\,$), now forbidden 
by $\,R$-invariance, was replaced by a trilinear coupling 
to an extra singlet chiral superfield $\,N\,$ 
transforming as 
$
\ \,N\,(\,x,\,\theta\,) \ \stackrel{R}{\longrightarrow}  \ \,
e^{2\, i\alpha}\ \,N\, (\,x, \,\theta \,e^{-i\alpha}\,)  \,, \
$ 
with an ``$R$-invariant'' superpotential written, 
in modern notations, as
$$
{\cal W}\ \ =\ \ \lambda\ \,H_1 H_2\,N\ +\ \sigma\ N\ \ .
$$
It has $\,R$-index $\,n=2\,$ and transforms according to eq.\,(\ref{rindex})
of section \ref{sec:rp}, so that it generates $\,R$-invariant 
interactions. Both the electroweak symmetry and the supersymmetry 
get spontaneously broken.
} \,(the discrete version of which leading to $\,R$-parity).
It also showed how one can break spontaneously 
the $\,SU(2) \times U(1)\,$ 
electroweak symmetry in a supersymmetric theory, 
using a pair of chiral doublet Higgs superfields 
that would now be called $\,H_1\,$ and $\,H_2\,$. This involves 
a mixing angle (initially called $\,\delta$) \,known as $\,\beta\,$,
\,defined by
\be
\tan \,\beta\ \ =\ \ \frac{v_2}{v_1}\ \ .
\ee
The fermions of this early supersymmetric model, 
which are in fact gaugino-higgsino mixtures,
should no longer be considered as lepton candidates, 
but became essentially the ``charginos'' and ``neutralinos'' 
of the Supersymmetric Standard Model~\cite{ssm,grav}.

\vskip.25truecm

Despite the general lack of similarities between known bosons and fermions, 
we tried as an exercise to see how far one could go in attempting 
to relate the spin-1 photon with a spin-$\frac{1}{2}$ neutrino.
If we want to attempt to identify the companion of the photon
as being a ``neutrino'',
despite the fact that it initially appears as a self-conjugate 
Majorana fermion, we need to understand 
how this particle could carry a conserved quantum number 
that we might interpret as a ``lepton'' number.
This was made possible through to the definition of
{{\it \,a continuous $\,U(1)\,$ $\,R$-invariance}}~\cite{R},
which also guaranteed the masslessness of this ``neutrino'' 
(``$\nu_L$'', ~carrying $\,+1\,$ unit of $\,R\,$),
\,by acting chirally on it, 
i.e. also by acting chirally on the Grassmann coordinate $\,\theta\,$
which appears in the expression of gauge and chiral superfields.
The supersymmetry generator $\,Q_\alpha\,$ carries one unit of the 
corresponding additive conserved 
quantum number, called $\,R,\,$
so that one has $\ \Delta \,R\,=\,\pm\,1$  between a boson 
and a fermion related by supersymmetry.

\vskip .25truecm
Attempting to relate the photon with one of the neutrinos
could only be an exercise of limited validity.
The would-be ``neutrino'', 
while having in this model a $\ V-A\ $ coupling to its associated ``lepton''
and the charged $\,W^\pm$ boson,
was in fact what we would now call a ``photino'', 
not directly coupled to the $\,Z\,$!
~Still this first attempt, which essentially became a part 
of the Supersymmetric Standard Model, 
illustrated how one can break spontaneously a
$\,SU(2) \times U(1)$ gauge symmetry in a supersymmetric theory, 
through an electroweak breaking induced by 
{\it a \underline{pair} \,of  chiral doublet Higgs superfields},
now known as $\,H_1\,$ and $\,H_2$ \,!
~(Using a single doublet Higgs superfield 
would have left us with a massless charged chiral fermion, which is
evidently unacceptable.)
~Our previous charged ``leptons''
were in fact what we now call two winos, or charginos,
obtained from the mixing of charged gaugino and higgsino components,
as given by the mass matrix

\vbox{\bea
\label{mwino}
{\cal M} =\hbox{\small $
\pmatrix{  (\,m_2\,=\,0\,) & \!\!\!
\displaystyle{\frac{g\,v_2}{\sqrt 2} = m_{W}\sqrt{2}\,\sin \beta }\,
\cr \cr
\displaystyle{\frac{g\,v_1}{\sqrt 2} = m_{W}\sqrt{2}\,\cos \beta } \!\!\!
&  \mu\,=\,0  \cr }\! ,
$}
\nonumber \\ [-.4truecm] \nonumber
\eea
\be
\ee
}
\noindent
in the absence of a direct higgsino mass that would have originated from a
$\ \mu\ H_1 H_2\ $ mass term 
in the superpotential\,\footnote{The $\ \mu\,H_1 H_2\,$ term
initially introduced in \cite{R},
which would have broken explicitly 
the continuous $\,U(1)\,$ $\,R$-invariance then intended to be
associated with ``lepton'' number conservation, 
was quickly replaced by a $\ \lambda\ \,H_1 H_2\,N\ $ trilinear 
coupling involving an 
{\it \,extra neutral singlet chiral superfield\,} $\,N\,$.
}.
The whole construction showed that one could deal elegantly 
with elementary spin-0 Higgs fields
(not a very popular ingredient at the time),
in the framework of spontaneously-broken supersymmetric theories.  
Quartic Higgs couplings are no longer arbitrary, 
but get fixed by the gauge coupling constants 
\,-- here the electroweak couplings 
\,\hbox{$g\,$ and $\,g'$ --\,}\, through the following ``$D$-terms''
\hbox{(i.e. 
\small $\frac{\vec D ^2}{2}\,+\,\frac{D'^2}{2}$
\normalsize
)}
in the scalar potential given in ~\cite{R}
\footnote{With a different denomination for the two Higgs doublets,
such that
$\ \varphi'' \ \mapsto \ h_1,$ \ \, $(\varphi')^c\ \mapsto
\ h_2,$ 
$\ \,\tan \delta = v'/v''\ \mapsto\ \tan \beta = v_2/v_1\ $.
}:
\bea
\ba{lcl}
\vspace{-.4cm}\\
V_{\hbox{\small{Higgs}}}\ \ =\ \   \vspace{4mm} \\
\hbox{\small $
\displaystyle{\ \frac{g^2}{8}\, 
(\,h_1^\dagger\vec \tau \,h_1 + h_2^\dagger\vec \tau \,h_2\,)^2 + 
 \frac{g'^2}{8}\,
(h_1^\dagger\,h_1-h_2^\dagger\,h_2)^2 +...}
$} 
\vspace{5mm}\\ 
 \displaystyle{ =\frac{g^2\!+g'^2}{8}\ 
(\,h_1^\dagger\,h_1-h_2^\dagger\,h_2\,)^2 +
\frac{g^2}{2}\,|\,h_1^\dagger\,h_2\,|^2+...\,. }
\ea
\nonumber
\eea
\be
\ee

\noindent
This is precisely the quartic Higgs
potential of the ``minimal'' version of the Supersymmetric Standard Model, 
the so-called MSSM, with its quartic Higgs coupling constants equal to
\be
\frac{g^2\,+\,g'^2}{8}\ \ \ \ \hbox{and}\ \ \ \ \ \frac{g^2}{2}\ \ .
\ee

\vskip .2truecm
The quartic term $\ \frac{g^2}{2}\ |\,h_1^\dagger\,h_2\,|^2\ $ 
in this scalar potential\,\footnote{For this quartic contribution 
to the Higgs potential, the correspondence 
between old and new notations is as follows:
$$
\frac{g^2}{2}\ \ (\varphi'^\dagger \varphi'\ \varphi''^\dagger \varphi''\ -\ 
\varphi'^\dagger \varphi''\ \varphi''^\dagger \varphi')\ \ \equiv\ \ 
\frac{g^2}{2}\ \ |\,h_1^\dagger\,h_2\,|^2\ \ .
$$
The vanishing of this quantity implies that the v.e.v.'s of the two Higgs 
doublets are correctly ``aligned'', with
$$
<h_1>\ \ =\,\ \left( \ba{c} v_1/\sqrt2 \vspace{1mm} \\ 0 \ea \right)\ , \ \ \ \
<h_2>\ \ =\,\ \left( \ba{c} 0 \vspace{1mm}\\ v_2/\sqrt2 \ea \right)\ .
$$
}
is responsible for the fact that 
the vacuum expectation values of the two doublet Higgs fields ``align'',
in order to minimize the energy,  
so that only neutral components of these doublets acquire 
non vanishing v.e.v.'s., the $\,U(1)$ subgroup 
of electromagnetism remaining unbroken, and the photon massless.
This term is also responsible, from $\,<\!h_{1,2}^{\ 0}\!>= v_{1,2}/\sqrt 2\,$, 
~for the contribution
$\ \frac{1}{4}\ g^2 (v_1^{\,2}+v_2^{\,2})\ |H^-|^2 \,=\, m_W^{\ 2}\ |H^-|^2$
to the mass$^2$ of the charged Higgs bosons $\,H^\pm\,$ 
(initially called $\,w^\pm$).

\vskip .2truecm

Further contributions to this quartic Higgs potential also appear
in the presence of additional superfields, such as the
neutral singlet chiral superfield $\,N\,$ already introduced in this model,
which will play an important r\^ole in the NMSSM, 
i.e. in ``next-to-minimal'' or ``non-minimal'' versions of 
the Supersymmetric Standard Model.
In any case charged Higgs bosons $H^\pm$ are present 
in this framework, as well as several neutral ones, now called $\,H,\ h,\ A,
\ ...$ , 
\,and one gets in general mass relations such as 
\be
m_{H^\pm}^{\ 2}  \ =\ m_W^{\ 2}\,+\, 
\hbox{\small susy-breaking terms},\ ...\ .
\ee
The exact mass spectrum depends of course on the details 
of the supersymmetry breaking mechanism considered:
use of soft-breaking terms, possibly ``derived from supergravity'', 
presence or absence of extra-$U(1)\,$ gauge fields 
and/or additional chiral superfields, r\^ole of radiative corrections, etc..

\vspace*{2mm}

\section{THE SUPERSYMMETRIC STANDARD MODEL}
\label{sec:ssm}

\vskip .05truecm
These two doublet Higgs superfields 
are precisely the two doublets, now called $\,H_1\,$ and $\,H_2\,$,
used in 1977 
to generate the masses of charged leptons
and down quarks (from $<\!H_1\!>$), and of up quarks (from $<\!H_2\!>$), 
in supersymmetric 
extensions of the standard model~\cite{ssm}. 
At the time 
having to introduce Higgs fields was generally 
considered as rather unpleasant.
While one Higgs doublet was taken as probably unavoidable 
to get to the standard model
or at least simulate the effects of the spontaneous breaking of the
electroweak symmetry, having to consider two Higgs doublets, 
necessitating charged Higgs bosons as well as several neutral ones,
was usually considered as a too heavy price, in addition to the 
``doubling of the number of particles'', once considered as
an indication of the irrelevance of supersymmetry.
As a matter of fact considerable work was devoted for a time on attempts 
to avoid fundamental spin-0 Higgs fields,
before returning to fundamental Higgses, 
precisely in this framework of supersymmetry.

\vskip .4truecm
In the previous $\,SU(2)\times U(1)\,$ model~\cite{R}, \,it was
impossible to view seriously for very long ``gaugino'' and ``higgsino'' fields 
as possible building blocks for our familiar lepton fields.
On the contrary they should describe new particles.
This led us to consider that all quarks and leptons
ought to be associated with new bosonic partners, the 
{\it spin-0 squarks and sleptons}.
Gauginos and higgsinos, mixed together 
by the spontaneous breaking of the electroweak symmetry,
correspond to a new class of fermions, now known as ``charginos''
(cf. the mass matrix of eq.\,(\ref{mwino}))
\,and ``neutralinos''.
In particular,
the partner of the photon under supersymmetry, which 
cannot be identified with any of the known neutrinos, 
should be viewed as a new ``photonic neutrino'',
the {\it \,photino\,};
the fermionic partner of the gluon octet is an octet of self-conjugate 
Majorana fermions called {\it \,gluinos\,}, 
etc. -- although at the time {\it \,colored fermions\,} belonging to 
{\it \,octet\,} 
representations of the color $\,SU(3)\,$ gauge group were generally believed 
not to exist (to the point that one could think of using the absence of 
such particles as a general constraint to select admissible 
grand-unified theories~\cite{gm}).

\vskip .3truecm

The two doublet Higgs superfields
$\,H_1$ and $\,H_2\,$ introduced previously can now be used to
generate quark and lepton masses\,\footnote{The correspondance 
between earlier notations 
for doublet Higgs superfields, and modern ones, is as follows:
$$
 \\ [.2 true cm]
\begin{tabular}{ccc}
$\!\!\!\!S\!=\!\left(\!\! \ba{cc}\! S^0 \vspace{.1truecm}\\ S^-\!
\ea \!\!\! \right)\! ,\, %\hbox{and}\ \ \,
T\!= \!\left(\!\! \ba{cc} \!T^0 \vspace{.1truecm}\\ T^-\!
\ea \!\!\!\right)\!\!\!\! $	 &  
$\!\!\rightarrow \!\! $   &  
$ \!\!\!\!
H_1\!=\!\left(\!\!\! \ba{cc} \!H_1^{\,0} \vspace{.1truecm}\\ H_1^{\,-}\!
\ea \!\!\!\right)\! ,\,
%\hbox{and}\ \ \,
H_2\!=\! \left( \!\!\!\ba{cc} H_2^{\,+} \vspace{.1truecm}\\ \!H_2^{\,0}
\ea \!\!\!\right)\!,$
                           \\  [-.1 true cm] && \\
(left-h.) \ \ \ \ (right-h.) &   &   (both left-h.) \ \ \ \ 
\end{tabular}
$$
so that $\,S\,$ gets now replaced by $\,H_1\,$ and $\,T^\dagger\,$ by $\,H_2$.
Furthermore, we originally denoted, generically, by $\,S_i$, \,left-handed, 
and $\,T_j$, \,right-handed,
the chiral superfields describing the left-handed and right-handed 
spin-$\frac{1}{2}$ quark and lepton fields, 
together with their spin-0 partners.}~\cite{ssm}.
To generate the appropriate mass terms we first write the relevant
bilinear products of left-handed lepton ($L$) and quark ($Q$) doublet 
superfields, with the conjugates of the right-handed superfields describing 
right-handed quarks and leptons, now known to be electroweak singlets. 
The latter are left-handed superfields 
denoted as $\,\bar E,\ \,\bar U$ and $\bar D$, \,in modern notations.
The resulting bilinear terms $\bar E\,L,\ \,\bar D \,Q$ and $\,\bar U\,Q\,$ 
are then coupled in a supersymmetric and gauge invariant way 
to the two doublet Higgs superfields $\,H_1\,$ and $\,H_2$, 
which leads to the familiar trilinear superpotential
\bea
\label{supot}
{\cal W} = \ h_e \ H_1 .\bar E \,L \,+\, 
h_d \ H_1. \bar D \,Q \,-\,  
h_u \  H_2 .\bar U \,Q \ .	
\nonumber		    
\eea
\be
\ee
The corresponding superpotential interactions are also invariant 
under continuous $\,R$-symmetry transformations, 
as well as under an ``extra-$U(1)$'' \,symmetry, 
two important features to which we shall return later.
We certainly don't attempt at this point to write further terms in 
the superpotential that would be {\it \,odd\,} functions 
of quark and lepton superfields, since we know in advance 
that they would lead to 
$\,B\,$ and/or $L\,$-violations, and it would be somewhat 
foolish to rush to reestablish the problems of  $\,B\,$ and $\,L\,$ 
conservation laws that were elegantly solved by attributing 
baryon and lepton numbers to bosons (squarks and sleptons) 
as well as to fermions.
(Such $\,B\,$ or $\,L$-violating terms, in addition, 
would not be invariant under $\,R\,$
nor under the extra $\,U(1)\,$ symmetry\,\footnote{It is interesting 
to note, in particular, that if the extra $\,U(1)\,$ of eq.\,(\ref{extra}) 
is gauged $\,B,\,L$ and $R\,$ become automatic symmetries which follow from 
the requirement of local gauge invariance. },
but we shall return later to the fate of these two additional symmetries.)

\vskip .3truecm

The vacuum expectation values of the two Higgs doublets 
generate charged-lepton and down-quark masses
(for the Higgs doublet described by $\,H_1$), 
and up-quark masses (for the one described by  $\,H_2$).
They are given, with an appropriate normalization convention, by
$\ m_e=h_e\,v_1/\sqrt 2\,,\ \,m_d=h_d\,v_1/\sqrt 2\,,$ ~and 
$\,m_u=h_u\,v_2/\sqrt 2\,$,
\,respectively.
All this constitutes the basic structure of the 
{\it \,Supersymmetric Standard Model\,},
which involves at least the minimal set of ingredients 
shown in Table \ref{tab:basic}.
Other ingredients, such as a $\,\mu\ H_1 H_2\,$ direct mass term 
in the superpotential, or an extra singlet chiral superfield $\,N\,$ 
with a trilinear superpotential coupling $\ \lambda \ H_1 H_2\,N\,+ \, ... \ $
possibly acting as a replacement for a  $\,\mu\ H_1H_2\,$ direct
mass term~\cite{R},
~and/or extra $\,U(1)\,$ factors in the gauge group
(which could have been responsible for spontaneous supersymmetry breaking)
may or may not be present, depending on the
particular version of the Supersymmetric Standard Model considered.

\vskip .3truecm

\begin{table}[t]
\caption{\ The basic ingredients of the Supersymmetric Standard Model.
\label{tab:basic}}
\vspace{0.2cm}
\begin{center}
\begin{tabular}{|l|} \hline \\ 
1) $\,SU(3)\times SU(2)\times U(1)\,$ gauge superfields; \\ \\
2) \,chiral superfields for  \\
\hskip 1truecm the three quark and lepton families; \\ \\
3) \,two doublet Higgs superfields $\,H_1$ and $\,H_2$
 \\ 
\hskip 1truecm responsible for electroweak breaking,  \\
\hskip 1truecm 
and quark and lepton masses, through 
%\hskip 1truecm through %the trilinear superpotential (\ref{supot}).
\\  [.3truecm]
4) \,the trilinear superpotential \,(\ref{supot})\ .
\\  \\ \hline
\end{tabular}
\end{center}
\end{table}

\begin{table}[t]
\caption{\ Minimal particle content of the Supersymmetric Standard Model.
\label{tab:SSM}}
\vspace{0.2cm}
\begin{center}
\begin{tabular}{|c|c|c|} \hline 
&&\\ [-0.2true cm]
Spin 1      
&\hbox{\hskip -1truecm Spin 1/2 \hskip -1truecm}    &Spin 0 \\ [.1 true cm]\hline 
&&\\ [-0.2true cm]
gluons  &\hbox{\hskip -1truecm gluinos ~$\tilde{g}$  \hskip -1truecm}      &\\
\,photon   \,     &\hbox{\hskip -1truecm photino ~$\tilde{\gamma}$ \hskip -1truecm } &\\ 
------------&$ - - - - - \, - $&-------------------- \\
 
%debut de la table intermediaire

$\ba{c}
W^\pm\\ [.1 true cm]Z \\ 
\\ \\
\ea $

&
$\ba{c}
\hbox {winos}  \ \widetilde W_{1,2}^{\,\pm} \\ 
[0 true cm]
\hbox {zinos } \ \ \widetilde Z_{1,2} \\ 
\\ 
\hbox {higgsino } \ \tilde h^0 
\ea$

& $\left.  \!\!\!\ba{c}
H^\pm\\
[0 true cm] H\ \\
\\
h, \ A
\ea\!\!\right\} 
\begin{array}{c} \hbox {Higgs}\!\!\!\\ \hbox {bosons}\!\!\! \end{array}$  \\ &&\\ 
[-.1true cm]
\hline &&

% fin de la table intermediaire
\\ [-0.2cm]
&leptons  ~$l$   \hskip -1truecm    &sleptons  ~$\tilde l$ \\
&quarks ~$q$       &squarks   ~$\tilde q$\\ [-0.3 cm]&&
\\ \hline
\end{tabular}
\end{center}
\end{table}

In any case, independently of the details of the
supersymmetry breaking mechanism 
ultimately considered, 
we obtain the minimal particle content
of the Supersymmetric Standard Model, as given in 
\hbox{Table \ref{tab:SSM}}.
Each \hbox{spin-$\frac{1}{2}$}\, quark $\,q\,$ or charged lepton $\,l^-\,$
is associated with {\it \,two\,} spin-0 partners collectively denoted by
$\,\tilde q\,$ or $\,\tilde l^-$, \,while a left-handed neutrino $\,\nu_L\,$ 
is associated with a {\it \,single\,} spin-0 sneutrino $\,\tilde \nu$.
\,We have ignored for simplicity 
further mixings between the various ``neutralinos''
described by neutral gaugino and higgsino fields, denoted in this table
by $\,\tilde\gamma,\ \tilde Z_{1,2}\,$ 
and $\tilde h^0$.
More precisely, all such models include four neutral Majorana fermions at least,
corresponding to mixings of the fermionic partners of
the two neutral $\,SU(2) \times U(1)\,$ gauge bosons (usually denoted by 
$\,\tilde Z\,$ and $\,\tilde\gamma$, 
~or $\,\tilde{W_3}\,$ and $\,\tilde B\,$) ~and of the 
two neutral higgsino components 
($\,\tilde{h_1^{\,0}}\,$ and $\,\tilde{h_2^{\,0}}$). 
\,Non-minimal models also involve additional 
gauginos and/or higssinos.

\vspace{3mm}

\section{FROM $R$-INVARIANCE TO $R$-PA\-RITY}
\label{sec:rp}

Let us  return to the definition of the continuous 
$R$-symmetry, and discrete $\,R$-parity, transformations.
$\,R$-parity is associated with 
a $\,Z_2\,$ subgroup of the group of continuous 
$\,U(1)$ $\,R$-symmetry transformations, acting on the gauge superfields 
and the two doublet Higgs superfields 
$\,H_1$ and $H_2\,$ as in \cite{R} 
(cf. footnote {\small \ref{footnote:R}} in section \ref{sec:R}),
with their definition extended to quark and lepton superfields
so that quarks and leptons carry $R=0\,$, 
~and squarks and sleptons, $\,R=\pm \,1\,$
(more precisely, 
$\,R=+\,1\, $ for $\,\tilde q_L,\,\tilde l_L$, ~and $\,R=-\,1\,$ 
for $\,\tilde q_R,\,\tilde l_R\,$)~\cite{ssm}.
$R$-parity appears in fact as the remnant of this continuous 
$\,R$-invariance when gravitational interactions are introduced~\cite{grav},
in the framework of local supersymmetry (supergravity).
Either the continuous $\,R$-invariance, 
or simply its discrete version of $\,R$-parity, if imposed, 
naturally forbid the unwanted direct exchanges of the new 
squarks and sleptons 
between ordinary quarks and leptons.

\vskip .3truecm

These continuous $\,U(1)$ $R$-symmetry transformations, 
which act chirally on the anticommuting Grassmann coordinate
$\,\theta\,$ appearing in the definition of superspace and superfields, act
on the gauge and chiral superfields of the Supersymmetric Standard Model
as follows\,\footnote{
If we also introduce as in \cite{R} an extra neutral singlet chiral 
superfield $\,N\,$ 
coupled to the two doublet Higgs superfields $\,H_1\,$ and $\,H_2$, 
\,it gets transformed under $\,R\,$ as follows:
$$
N (\,x,\,\theta\,)  \ \ \stackrel{R}{\longrightarrow}  \ \,e^{2\,i\alpha}  \ \
N (\,x, \,\theta \,e^{-i\alpha}\,) \ \ ,
$$
so that the trilinear superpotential coupling
$\,\lambda\ H_1 H_2\,N\,$ is \,``$R$-invariant'', in the sense 
that it transforms according to eq.\,(\ref{rindex}).
}:

\bea
\label{r}
\ba{c}
\!\!\!\!\!
\hbox{\underline{Action of continuous $\,R$-symmetry}\,:} 
\nonumber \\ [.5truecm]
\left\{ \ 
\ba{lcl}
\ V(\,x,\,\theta,\,\bar\theta\ ) &\ \,\stackrel{R}{\longrightarrow} \,\ &  
V (\,x, \,\theta \,e^{-i\alpha},\,\bar\theta \, e^{i\alpha}\,) \ ,  
\\ [.3truecm]
\ H_{1,2} \,(\,x,\,\theta\,) &\stackrel{R}{\longrightarrow}  &
H_{1,2}\, (\,x, \,\theta \,e^{-i\alpha}\,)  \ \ ,  
 \\[.3truecm]
\ S (\,x,\,\theta\,) &\stackrel{R}{\longrightarrow}  & e^{i\alpha}  \ \
S (\,x, \,\theta \,e^{-i\alpha}\,) \ \ , 
\ea  
\right.  
\ea  
\nonumber
\eea
\bea
\hbox{for}\ \left\{ \ba{l}
\hbox{ $\!SU(3)\!\times\! SU(2)\!\times \!U(1)\,$ gauge superfields,}  
\\ [.2truecm]
\hbox{left-h. Higgs superfields
$\,H_1\,$ and  $\,H_2\,,$}  \\[.2truecm]
\hbox{left-h.
(anti)quark and lepton superfields} \\ [.1truecm]
   \hskip 1truecm \hbox{\small $ \ \ S\ \ =\ \ \{\ Q, \,\bar U, \,\bar D,\ \,
L, \,\bar E\ \}$} \ \ ,
\ea \right.
\nonumber
\eea

\be
\hskip 4truecm  \hbox{respectively}\ \ .
\ee

\vskip .1truecm

\noindent
These transformations are defined so as not to act on ordinary particles, 
which have $\,R=0\,$, 
\,while their superpartners have, therefore, $\,R=\pm1\,$.
\,They allow us to distinguish between two separate sectors
of $\,R$-even and $\,R$-odd particles. 
$\,R$-even particles 
%(having $\,R$-parity $\,R_p\,=\,(-1)^R\,=\,+\,1\,$) 
include the gluons, photon,
$\,W^\pm$ and $\,Z$ bosons, the various Higgs bosons, 
quarks and leptons 
\,-- and the graviton.
$\,R$-odd particles 
%(having $\,R$-parity $\,R_p\,=\,(-1)^R\,=\,-\,1\,$) 
include their superpartners, i.e. the gluinos 
and the various neutralinos and charginos, squarks and sleptons
\,-- and the gravitino (cf. Table \ref{tab:Rp}).
According to this first definition, $\,R$-parity simply appears as
the parity of the above additive quantum number $\,R\,$,
as given by the expression~\cite{rp}:
\bea
\label{rp01}\ba{l} \\ [-.3truecm]
R\hbox{-parity}\ \ R_p\ =\ 
\\  [.2truecm]
\ \ \ (\,-\,1\,)^{R}\ \ =\ \ \left\{ 
\ba{l} 
+\,1\ \ \ \hbox{for ordinary particles,} \vspace{2mm} \\
-\,1\ \ \ \ \hbox{for superpartners.}
\ea  \right. \\ [-.1truecm]
\ea
\nonumber
\eea
\be
\ee

\vskip .1truecm
But why should we limit ourselves to the discrete $\,R$-parity
symmetry, rather than considering its full continuous parent
$\,R$-invariance\,?
This {\it \,continuous\,} $\,U(1)\,$ $\,R$-invariance, from which 
$\,R$-parity has emerged, is indeed a symmetry of all
four necessary basic building blocks 
of the Supersymmetric Standard Model~\cite{ssm}:

\vskip .2truecm
1) the Lagrangian density 
for $\,SU(3)\times SU(2) \times U(1)\,$ gauge superfields;

\vskip .15truecm

2) the $\,SU(3)\times SU(2) \times U(1)\,$ gauge interactions 
of the quark and lepton superfields;

\vskip .15truecm

3) the $\,SU(2) \times U(1)\,$ gauge interactions 
of the two chiral doublet Higgs superfields $\,H_1\,$ and $\,H_2\,$
responsible for the electroweak breaking;

\vskip .15truecm

4) and the trilinear ``superYukawa'' interactions (\ref{supot})
responsible for quark and lepton masses.

\vskip .4truecm

Indeed the bilinear products of left-handed (anti)lepton 
and (anti)quark superfields 
$\,\bar E \,L$, $\bar D \,Q \,$ and $\,\bar U \,Q\ $ 
that we have to consider to generate quark and lepton mass terms in a 
supersymmetric extension of the standard model 
\,transform under $\,R\,$ as follows, according to eq.\,(\ref{r}):
\be
\bar E \,L \,(\,x,\,\theta\,) \ \,\stackrel{R}{\longrightarrow}  \ \,
e^{2\,i\alpha} \ \,
\bar E \,L \,(\,x, \,\theta \,e^{-i\alpha}\,) \,,\ ...\,.
\ee
When these terms are combined with $\,H_1\,$ 
(for $\bar E \,L\,$ and $\,\bar D \,Q \,$) and  $\,H_2\,$ 
(for $\,\bar U \,Q \,$) to give the trilinear superpotential (\ref{supot}), 
this one transforms under the continuous 
$\,R$-symmetry (\ref{r}) with 
``$\,R$-weight'' $\ n_{\cal W}\,=\,\sum_i\,n_i\,=\,2\,$,
~i.e. according to

\be
\label{rindex}
{\cal W}\,(\,x,\,\theta\,) \ \ \stackrel{R}{\longrightarrow}\ \  e^{2\,i\alpha}\ \ 
{\cal W}\,(\,x, \,\theta \,e^{-i\alpha}\,)\ \ .
\ee

\vskip .2truecm
\noindent
Its auxiliary ``$\,F$-component'' (obtained from the coefficient of the 
bilinear $\,\theta \,\theta \,$ term in the expansion of
$\,{\cal W}\,)$, \,is therefore $\,R$-invariant, 
generating $\,R$-inva\-riant interaction terms 
in the Lagrangian density\,\footnote{Note, however, that a direct Higgs 
superfield mass term
$\,\mu\ H_1 H_2\,$ in the superpotential, which has $\,R$-weight $\,n=0\,$,
\,does {\it \,not\,} lead to interactions which are invariant 
under the continous $\,R\,$ symmetry; but it gets in general reallowed, as 
for example in the MSSM, 
as soon as the continuous $\,R$ symmetry gets 
reduced to its discrete version of $\,R$-parity.}.

\begin{table}[t]
\caption{\ $R$-parities 
in the Supersymmetric Standard Model.
\label{tab:Rp}}
\vspace{0.2cm}
\small
\begin{center}
\begin{tabular}{|c|c|} \hline 
&\\ [-0.2true cm]
{\normalsize Bosons}  & {\normalsize Fermions} \\ [.2 true cm]\hline \hline 
&\\ [-0.1true cm]
$\!\!\!\! 
\ba{c}
\hbox{gauge and Higgs bosons}  \\ [.1 true cm]
\hbox{graviton}
\ea \!\!\!\!
$
& 
$\!\!\!\!
\ba{c}
\hbox{gauginos and higgsinos}  \\ [.1 true cm]
\hbox{gravitino}
\ea\!\!\!\!
$
\\  [0.1 cm]& \\
$(\,R\,=\,0\,)$  & $(\,R\,=\,\pm 1\,)$
\\ & \\ 
{\normalsize  $R$-parity\ \ $+$}  & {\normalsize $R$-parity\ \ $-$}
\\ & \\ \hline & \\[-0.1cm]
sleptons and squarks & 
leptons and quarks \\ [0.1 cm] & \\
(\,$R\,=\,\pm\,1$\,)  & $(\,R\,=\,0\,)$ \\ & \\
{\normalsize $R$-parity\ \ $-$}  & {\normalsize $R$-parity\ \ $+$} 
\\ & \\ \hline
\end{tabular}
\end{center}
\end{table}
\normalsize

\vskip .3truecm
However, an unbroken continuous $\,R$-invariance,
which acts chirally 
on the Majorana octet of gluinos,
\be
\tilde g\ \ \stackrel{R}{\longrightarrow}\ \ e^{\,\gamma_5\,\alpha}\ \tilde g\ \ .
\ee
would constrain gluinos
to remain massless, even after a breaking of the supersymmetry.
We would then expect the existence of relatively light
``$R$-hadrons''~\cite{ff,ff2} made of quarks, antiquarks and gluinos, 
which have not been observed. We know today that gluinos, 
if they do exist, should be rather heavy, requiring a significant 
breaking of the continuous $\,R$-invariance,
in addition to the necessary breaking of the supersymmetry.
Once the continuous $\,R$-invariance is abandoned, 
and supersymmetry is spontaneously broken,
radiative corrections do indeed allow for the generation of 
gluino masses~\cite{glu}, a point to which we shall return later
(cf. section \ref{sec:grav}).

\vskip .4truecm
Furthermore, the necessity of generating
a mass for the Majorana spin-$\frac{3}{2}\,$ {\it gravitino}, 
\,once {\it \,local\,} supersymmetry 
is spontaneously broken, also forces us to abandon 
the continuous $\,R$-invariance, 
in favor of the discrete $\,R$-parity symmetry, thereby 
also allowing for gluino
and other gaugino masses, at the same time as the gravitino mass $\,m_{3/2}$, 
\,as already noted in 1977~\cite{grav}.
A third reason for abandoning the continuous $\,R$-symmetry 
could now be the non-observation at LEP
of a charged {\it wino} \,-- also called {\it \,chargino\,} --\,
lighter than the $\,W^\pm$, \,that would exist 
in the case of a continuous $\,U(1)\,$ $\,R$-invariance~\cite{R,ssm},
\,as shown by the mass matrix $\,{\cal M}\,$ of eq.\,(\ref{mwino})
(the just-discovered $\,\tau^-\,$ particle could tentatively be considered, 
in 1976, as a possible light wino/chargino candidate, 
before getting clearly identified as a sequential heavy lepton.)

\vskip .5truecm
Once we drop the continuous $\,R$-invariance 
in favor of its discrete $\,R$-parity version,
we may ask how general is this notion of $\,R$-parity, and,
correlatively, are we {\it \,forced\,}
to have this $\,R$-parity conserved\,?
As a matter of fact,
there is from the beginning a close connection between $\,R$-parity 
and baryon and lepton number conservation laws, 
which has its origin in our desire 
to get supersymmetric theories in which $\,B\,$ and $\,L\,$ 
could be conserved, and, at the same time, 
to avoid unwanted exchanges of spin-0 squarks and sleptons.
Actually the superpotential of the theories discussed in Ref.\,\cite{ssm}
\,was constrained from the beginning, for that purpose,
to be an {\it \,even\,} function of quark and lepton superfields,
more specifically involving, for left-handed electroweak doublets 
and right-handed singlets, the bilinear combinations 
$\,\bar E L,\ \bar D Q\,$ and $\bar U Q$.
{\it Odd\,} superpotential terms,
which would have violated the ``matter-parity'' symmetry 
$\,(\,-1)^{(3B+L)}$, ~were excluded,
to be able to recover $\,B\,$ and $\,L\,$ conservation laws,
and avoid direct Yukawa exchanges of spin-0 squarks and sleptons
between ordinary quarks and leptons.
Tolerating unnecessary superpotential terms which are {\it odd\,} 
functions of quark and lepton superfields
\,(i.e. $\!\!R_p$-violating terms), 
does create, in general, immediate problems
with baryon and lepton number conservation laws
(most notably, a much too fast proton instability, if both 
$\,B\,$ and $\,L\,$ violations are simultaneously allowed).

\vskip .25truecm
This intimate connection between $R$-parity 
and  $B\,$ and $\,L\,$ conservation laws can be made quite obvious
by noting that for usual particles $\,(-1)\,^{2S}$ 
coincides with $\, \ (-1)\,^{3B+L}$, \,so that the original
$R$-parity (\ref{rp01}) may be reexpressed 
in terms of the spin $\,S\,$ and the ``matter-parity'' $(-1)\,^{3B+L}\,$, 
~as follows~\cite{ff}:
\vskip -.55truecm
\bea
\label{rp02}
R\hbox{-parity} \ \ =\ \ (-1)\,^{2S} \ (-1)\,^{3B+L}    \ \ .     
\eea

%\vskip .2truecm
\noindent
This may also be written as $\ (-1)^{2S} \ (-1)\,^{3\,(B-L)}\,$, 
~showing that this discrete symmetry may still be conserved 
even if baryon and lepton numbers are separately violated,
as long as their difference ($\,B-L\,$) remains 
conserved, at least modulo 2.

\vskip .35truecm

The $\,R$-parity symmetry operator
may also be viewed as a non-trivial geometrical discrete symmetry 
associated with 
a reflection of the anticommuting fermionic Grassmann coordinate, 
$\,\theta\ \to -\,\theta\,$, in superspace~\cite{geom}.
This $\,R$-parity operator plays an essential r\^ole 
in the discussion of the experimental signatures of the new particles 
(even if it should turn out not to be exactly conserved).
A conserved $\,R$-parity guarantees 
that {\it \,the new spin-0 squarks and sleptons 
cannot be directly exchanged\ } between ordinary quarks and leptons, 
as well as the absolute stability of the ``lightest supersymmetric particle'' 
(or LSP), a good candidate for non-baryonic Dark Matter 
in the Universe.

\vspace{2mm}

\section{ABOUT SUPERSYMMETRY BREA\-K\-ING, IN THE EARLY TIMES}
\label{sec:grav}

%\vskip .05truecm 

Let us come back to the question of supersymmetry breaking,
which still has not received a definitive answer yet.
The inclusion, in the Lagrangian density, 
of universal soft dimen\-sion-2\, supersymmetry breaking terms 
for all squarks and sleptons,
\be
\label{soft}
{\cal L}_{\hbox{\footnotesize{soft}}}\ \ =\ \ 
-\ \sum_{\tilde q,\,\tilde l}\ m_0^{\,2}\ \ 
(\,{\tilde q}^\dagger\,\tilde q\ +\ {\tilde l}^\dagger \,\tilde l\,)\ \ ,
\ee
was already considered in 1976.
Such terms breaking explicitly the supersymmetry, 
as soon as they are allowed or tolerated,
can immediately make superpartners heavy with no difficulty, 
in supersymmetric extensions of the standard model
with an $SU(3)\times SU(2)\times U(1)$ gauge group
\,(no extra $U(1)$\,), 
\,and quarks and charged leptons massive through their couplings to 
$H_1$ and $H_2$ \cite{ssm}.
These models may involve an additional singlet superfield $N$
coupled to $H_1$ and $H_2$ as in \cite{R},
with a trilinear coupling $\ \lambda\ H_1H_2\,N\,+\,...\, $ 
in place of the initial $\,\mu\,H_1H_2$ (cf. ``NMSSM'').
One may also disregard this extra superfield, 
then restoring the $\,\mu\,H_1H_2\,$ mass term,
which leads to the minimal version of the Supersymmetric 
Standard Model\,\footnote{In both cases, the photon and gluons remain massless
as they should,
and there is no massless or light 
\hbox{spin-0} boson since no spontaneously broken extra $\,U(1)\,$  
is present, once the singlet $\,N\,$ and/or the various 
soft-breaking terms are included.}.

\vskip .3truecm 
We were however more ambitious, since it was also understood 
that such soft-breaking terms of dimension 2 should better originate from
a spontaneous supersymmetry breaking mechanism, 
especially if supersymmetry is to be realized locally. 
This required additional work.
As a matter of fact, in view of progressing towards a true spontaneous 
breaking mechanism,
the soft breaking terms of eq.\,({\ref{soft}) 
were first generated spontaneously 
with the help of the ``$D$-term'' associated with 
an \,``{\it extra} $\,U(1)$''\, gauge symmetry, 
acting {\it \,axially\,} on lepton and quark fields, thereby 
providing a common positive mass$^2$ contribution $\,m_0^{\,2}\,$
(that was initially called $\,\mu^2$)
for all (``left-handed'' as well as ``right-handed'') 
slepton and squark fields. 
One can then disregard this potentially unpleasant extra $\,U(1)\,$ 
by sending it to an ``invisible sector'', ultimately taking 
the limit in which its gauge coupling $\,g''\,$ vanishes, 
as done in the first paper of ref.\,\cite{ssm}.
The spin-$\frac{1}{2}\,$ goldstino (here 
the gaugino of the extra $\,U(1)\,$) \,gets ``invisible'' 
then decouples. The supersymmetry, first spontaneously broken 
but ``at a very high scale'' 
(here $\,\sqrt d = \Lambda_{ss} \approx
\sqrt {\, m_0^{\,2}/g''}\, \gg m_W$),    
\,remains broken \hbox{-- now explicitly --\,}  in the limit;
but only softly, through the slepton and squark (universal, 
in the simplest case) dimension-2 \,mass terms (\ref{soft}).
%(This method of generating soft generating supersymmetry breaking terms 
%from a theory in which supersymmetry is spontaneously broken 
%``at a high scale'' so that the Goldstone spinor almost decouples 
%is somewhat comparable to the ``gravity-induced'' breaking mechanisms 
%that were discussed later.)
The approach illustrated also very well the special r\^ole played by the 
``susy-breaking scale parameter'', which could well be very large 
while the mass splittings between bosons and fermions still 
remain much smaller, typically $\,\simle\,$ electroweak scale.

\vskip .3truecm

Considering such an extra $\,U(1)\,$ is indeed natural 
in the framework of supersymmetric theories, in 
which two doublet Higgs superfields are separately responsible 
for the masses of charged leptons and down quarks ($H_1$), 
and of up quarks ($H_2$), as we have seen.
One may therefore take advantage of this special 
feature of supersymmetric theories, 
also imposed by the necessity of avoiding massless charged fermions,
to perform independent phase transformations on the two Higgs doublets.
More precisely one can perform
$\,U(1)_Y \times \hbox{extra-}U(1)\,$
gauge transformations, 
acting independently on the two Higgs doublets $\,H_1$ and $\,H_2\,$.
This corresponds to the invariance of the trilinear 
superpotential (\ref{supot}) responsible for quark and lepton masses
under an extra $\,U(1)\,$ symmetry defined, in the simplest case, 
as follows:
\bea
\label{extra}
\ba{c}
\!
\hbox{{Action of extra $\,U(1)$}\,:} 
\nonumber \\ [.4truecm]  
\!
\!\!\left\{ \ 
\ba{ccccl}
V(\,x,\,\theta,\,\bar\theta\ )&
%\stackrel{\hbox{\small extra $U(1)$}}
{\longrightarrow} &  
V(\,x,\,\theta,\,\bar\theta\ ) \ \ ,
\\ [.25truecm]
H_{1,2} (x,\,\theta) & \longrightarrow  &
\ e^{-\,i\alpha}\ H_{1,2} (x,\,\theta) \ \, , 
\\  [.25truecm]
S (x,\,\theta) &\ \ \longrightarrow \ \  & e^{\,i\frac{\alpha}{2}}  \ 
S (x,\,\theta)\ \ , \ea
\right. 
\ea
\nonumber \\   \nonumber
\eea
\bea
\ba{c}
\!\!\!
\hbox{for}\ \left\{ \ba{l}
\hbox{ $\!SU(3)\!\times\! SU(2)\!\times \!U(1)\,$ gauge superfields,}  
\\ [.2truecm]
\hbox{left-h. Higgs superfields
$\,H_1\,$ and  $\,H_2\,,$}  \\[.2truecm]
\hbox{left-h.
(anti)quark and lepton superfields} \\ [.1truecm]
   \hskip 1truecm \hbox{\small $ \ \ S\ \ =\ \ \{\ Q, \,\bar U, \,\bar D,\ \,
L, \,\bar E\ \}$} \ \ .
\ea \right.
\nonumber
\ea
\nonumber
\eea

\be
\ee

\noindent
This extra $\,U(1)\,$ is associated, in the above simple case,
with a purely axial new neutral current 
for all quarks and charged leptons\,\footnote{
When we also introduce the extra neutral singlet chiral superfield $\,N\,$ 
coupled to $\,H_1\,$ and $\,H_2$, 
it gets transformed under the extra $\,U(1)\,$ as follows:
$$
N (\,x,\,\theta\,)  \ \ \longrightarrow  \ \  e^{2\,i\alpha}  \ \
N (\,x, \,\theta\,)
$$
so that its trilinear superpotential coupling 
$\,\lambda\ H_1 H_2\,N$ 
be invariant under the extra $\,U(1)$. 
(A superpotential term linear in $\,N$,
\,if present as in \cite{R}, 
would however break explicitly the extra $\,U(1)\,$ symmetry, 
as would also do $\,N^2$ or $\,N^3\,$ terms.)
}.
The whole construction is designed so as to make the set of equations
$\ <\!D\!>'\hbox{s}\,=\,\,<\!F\!>'\hbox{s}\,=\,\hbox{$<\!G\!>'\hbox{s}$}
\,=\,0$  to have no solution at all, in order to
 obtain a true spontaneous breaking of the supersymmetry 
with a physically coupled goldstino (later to be ``eaten'' by the gravitino),
as well a spontaneous breaking of the extra $\,U(1)$ 
and of the electroweak symmetry, with the $\,SU(3)$$\times $$U(1)\,$ 
of QCD $\times $ QED remaining unbroken: quite a non-trivial result given 
all the constraints to be satisfied simultaneously.
The soft breaking terms $\,m_0^{\,2}\,$ of eq.\,(\ref{soft}) 
are then generated by
the v.e.v. of the $D$-component associated with the extra $\,U(1)\,$, 
which reads in the simplest case:
\be
m_0^2\ \ =\ \ \frac{1}{2}\ \ g''\ <\! -D''\!>\ \ ,
\ee
so that one gets the mass relations:
\bea
\label{spec}
\left\{\ \ba{ccc}
m_{\tilde q}^{\,2} &=& m_q^{\,2}\ +\ m_0^2 \ \ , \\ [.2truecm]
m_{\tilde l}^{\,2} &=& m_l^{\,2}\ +\ m_0^2  \ \ .
\ea \right.
\eea
Additional contributions to $\,m_0^{\,2}\,$, which depend linearly on the weak 
isospin and hypercharge quantum numbers $\,T_3\,$ and $\,Y$,
\,may also be generated from non-vanishing 
contributions of the $<\!\!D\!\!>\,$'s 
\,associated with $SU(2)\times U(1)\,$, 
but they remain family-independent (which is useful to avoid 
potential difficulties with flavor-changing neutral current effects).
In the presence of such terms, the previous formula (\ref{spec}) becomes:
\bea
\label{delta}
m_{\tilde q,\tilde l}^{\ 2} \!\!&=& \!\!m_{q,l}^{\,2}\ +\ m_0^2\ 
\nonumber\\ [.1truecm] 
&&\mp \ \hbox{\small $\left(\, 
g\ T_3\,<\!D_3\!>+\,\frac{g'}{2}\ Y \,<\!D'\!>\,\right)$ } ,\,
\eea
in which the sign $\,\mp\,$ corresponds to the two different (left and right,
respectively)
handedness of the sfermion fields considered.

\vskip .3truecm

This illustrates clearly {\it the reason for 
 having introduced such an extra $\boldmath U(1)$},
\,in order 
to get spontaneous supersymmetry breaking with heavy squarks and sleptons.
Indeed 
in a spontaneously broken globally supersymmetric theory with 
$\,SU(3)\times SU(2)\times U(1)\,$ (or a fortiori $\,SU(5)\,$) 
as the gauge group, 
possible supersymmetry-breaking contributions 
to the mass$^2$ matrices of squarks and sleptons
originating from the weak hypercharge and weak isospin
``$D$-terms'' in the scalar potential always depend linearly on the
$\,Y\,$ and $\,T_3\,$ gauge quantum numbers of 
(left-handed) matter superfields, and therefore cannot all be positive;
and $\,F$-term contributions mixing squarks or sleptons together 
cannot come to the rescue.
As a result one at least among the four $\,\tilde u\,$ and 
$\,\tilde d\,$ squarks would then have a negative mass$^2$
(unless they all remain light), which is evidently unacceptable.
This is well illustrated by the mass sum rule established in \cite{extrau},
which reads, {\it \,in the absence of such an extra} $\,U(1)\,$:
\be
\sum \ m^2(\hbox{squarks})\ \ = \ \ 2\ \ \sum \ m^2(\hbox{quarks}) \ .
\ee
The average mass$^2$ is the same for squarks and quarks, for example
(up to radiative corrections),
in a spontaneously broken globally supersymmetric theory,
in the absence of an extra $\,U(1)$. 
The use of a ``non-traceless'' extra $\,U(1)\,$, as we had introduced 
a few years before precisely for that reason, 
allowed for a non-vanishing supertrace: 
\bea
\nonumber \\ [-.3truecm]
\ba{l}
\!\!\!\hbox{Supertrace} \ {\cal M}^2_{\hbox{\small{\,(squarks and quarks)}}}
\\ \\
\!=\ \displaystyle{\sum \ m^2(\hbox{\small squarks})\ \ -\ \ 
2\ \sum \ m^2(\hbox{\small quarks}) \ \ }
\\ [.3truecm]
\!\approx \ g''\,< \!D_{\hbox{\scriptsize{extra-$U(1)$}}}\!\! > \ 
\hbox{Tr}_{\hbox{\scriptsize{quark sf.}}} \,\hbox{\small extra-}U(1)\,
{\bf  > \, 0\,};
\ea
\nonumber
\eea
\be
\ee

\noindent
and for heavy squarks and sleptons, already at the classical level, 
in spontaneously broken global supersymmetry.

\vskip .3truecm
The problem of having all squarks and sleptons heavy, 
in a spontaneously broken supersymmetric theory, 
appears solved for the moment, at the price of the 
extension of the gauge group to include an extra $\,U(1)\,$ factor.
Otherwise one would have to rely, as it was proposed later, 
on rather complicated effects of radiative corrections, 
or to go to supergravity theories.
Of course, in all cases one may always decide to return 
to explicit supersymmetry breaking,
which immediately leads us to physically acceptable 
$\,SU(3)\times SU(2)\times U(1)\,$ theories, without or with 
additional neutral singlet chiral superfields. 
This makes the difficult question of supersymmetry breaking 
immediately disappear, 
although one still has to wonder about the physical origin of this breaking.

\vskip .25truecm

The previous method of spontaneous supersymmetry breaking
quickly faced several difficulties. 
In addition to the question of anomalies 
(connected with the non-traceless feature of the extra $\,U(1)$,
but maybe anomalous $\,U(1)$'s could be tolerated after all ...\,),
it required new neutral current interactions.
This was fine in 1977, but such interactions
did not show up, as the $SU(2) \times U(1)\,$ neutral current structure 
of the Standard Model got brilliantly confirmed. 
(The question is in fact more complex than it seems, in particular since
one may also consider situations for which the new gauge boson 
would be both very light and very weakly coupled, but we do not elaborate 
on this here\,.)
This mechanism also 
left us with the question of generating large gluino masses, 
to which we shall return soon.
Altogether, the gauging of an extra $U(1)\,$ no longer appears
as an appropriate way to generate large superpartner masses \,-- 
even if such an $U(1)$ may still have a r\^ole to play --\,
and we can now close this parenthesis.
One now uses again, in general, soft supersymmetry-breaking 
terms~\cite{gg} generalizing those of eq.\,(\ref{soft}) 
\,-- possibly ``induced by supergravity'', or other mechanisms 
for which the supersymmetry is in general spontaneously broken 
``at a high scale''.
These terms essentially serve as a parametrization 
of our ignorance about the 
true mechanism of supersymmetry breaking chosen by Nature
to make superpartners heavy.

\vskip .36truecm

But let us return to \underline{\bf gluino masses}. As we said before 
continuous $\,R$-symmetry transformations act {\it \,chirally\,} 
on gluinos, so that an unbroken $\,R$-invariance
would require them to remain massless,
even after a spontaneous breaking of the supersymmetry\,!
Thus the need, once it became experimentally 
clear that massless or even light gluinos could not be tolerated, 
to generate a gluino mass either from radiative
corrections~\cite{glu}, or from supergravity (see already {\cite{grav}), 
with, in both cases, the continuous $\,R$-invariance reduced to its 
discrete $\,R$-parity subgroup.

\vskip .3truecm
In the framework of global supersymmetry
it is not so easy to generate large gluino masses.
Even if global supersymmetry is spontaneously broken, 
and if the continuous $R$-symmetry is not present, 
it is still in general rather difficult to obtain large masses for gluinos, 
since: \ 
{\bf i)} \ no direct gluino mass term is present in the Lagrangian density; 
and \ 
{\bf ii)} \ no such term may be generated spontaneously, at the tree 
approximation, since gluino couplings involve {\it colored\,} 
\hbox{spin-0} fields, which cannot be translated. 
A gluino mass may then be generated by radiative corrections
involving a new sector of quarks sensitive 
to the source of supersymmetry breaking~\cite{glu},
that would now be called  ``messenger quarks'',
but \ {\bf iii)} \ this can only be through diagrams which ``know'' both about:
\, {\bf a)} \, the spontaneous breaking of the global supersymmetry,
through some appropriately-generated v.e.v.'s for auxiliary components,
$\,<\!D\!>,\ <\!F\!>\,$ or $\,<\!G\!>\,$'s;\ \ 
\, {\bf b)} \, the existence of superpotential interactions 
which do not preserve 
the continuous $\,U(1)\,$ $R$-symmetry.
%Ref. \cite{glu} showed that it was indeed possible to generate 
%gluino masses by radiative corrections, 
%through the interaction of gluinos with an ``ad hoc'' 
%sector of what would be called now {\it \,vectorlike ``messenger'' quarks\,}, 
%sensitive to the spontaneous breaking of the supersymmetry.
Such radiatively-generated gluino masses, however, 
generally tend to be rather small, unless one introduces, in 
some often rather complicated ``hidden sector'', 
very large mass scales $\,\gg\,m_W\,$.

\vskip .45truecm

     Fortunately {\bf gluino masses} may also result directly from {\bf 
\underline{supergravity}}, as already observed in 1977~\cite{grav}. 
Gravitational 
interactions require, within local supersymmetry, that the spin-2 
graviton be associated with a spin-$3/2\,$ partner~\cite{sugra}, the 
gravitino. Since the gravitino is the fermionic gauge particle of 
supersymmetry it must acquire a mass, $\,m_{3/2} \ (= \kappa \ d/ \sqrt{6}
 \ \approx \,
d/m_{\rm{Planck}}\,\,)$, as soon as the local supersymmetry gets spontaneously 
broken. 
Since the gravitino is a self-conjugate Majorana fermion
its mass breaks the continuous $\,R$-invariance which acts chirally on it,
just as for the gluinos, 
\,forcing us to abandon the continuous $U(1)$ $R$-invariance, 
in favor of its discrete $R$-parity subgroup.
In particular, in the presence of a spin-$\frac{3}{2}\,$ gravitino mass 
term $\,m_{3/2}\,$,
~which corresponds to a change in $\,R\,$ $\ \Delta R=\pm \,2\,$,
~the ``left-handed sfermions''
$\,\tilde f_L$, ~which carry $\,R=+\,1$, ~can mix with the 
``right-handed'' ones  $\,\tilde f_R$, 
~carrying $\,R=-\,1$, ~through mixing terms having 
$\ \Delta R=\pm \,2\,$, ~which may naturally \,(but not necessarily)
be of order $\ m_{3/2}\ m_f\,$ -- so that the lightest of the sfermions 
may well turn out to be one of the two stop quarks $\,\tilde t\,$.
\linebreak
Supergravity theories also offer, in addition, a natural framework 
in which to include direct gaugino Majorana mass terms 
\bea
\nonumber \\ [-.4truecm]
-\,\frac{i}{2}\ m_3\ \,\bar {\tilde G}_a\,\tilde G_a\ 
-\,\frac{i}{2}\ m_2\ \,\bar {\tilde W}_a\,\tilde W_a\ 
-\,\frac{i}{2}\ m_1\ \,\bar {\tilde B}\,\tilde B\,,
\nonumber 
\eea
\be
\ee
which also correspond to $\,\Delta R=\pm \,2\,$,
\,just as for the gravitino mass itself.
$\!$The $\,SU(3) \times SU(2) \times U(1)\,$ gaugino 
mass parameters $m_3,\ m_2\,$ and $\,m_1\,$
may naturally \,(but not necessarily)\,
be of the same order as the gravitino mass $\,m_{3/2}\,$.

\vskip .45truecm

Once the continuous $R$-invariance 
is reduced to its discrete $R$-parity subgroup, 
a direct Higgs superfield mass term $\ \mu \ H_1 H_2$,
~which was not allowed by the continuous $U(1)\,$
$R$-symmetry, gets reallowed in the superpotential,
as for example in the MSSM. The size of this 
supersymmetric $\,\mu\,$ parameter might conceivably have been 
a source of difficulty, in case this parameter, present even 
if there is no supersymmetry breaking, turned out to be large.
But since the $\mu$ term breaks explicitly 
both the continuous $\,R$-invariance (\ref{r})
and the (global) extra $\,U(1)\,$ symmetry (\ref{extra}) its size
may be controlled by considering one or the other of these two symmetries.
Even better, since $\mu$ got reallowed just as we abandoned 
the continuous $\,R$-invariance so as to allow for gluino and gravitino masses,
the size of $\,\mu\,$ may naturally be 
of the same order as these gaugino (and gravitino) masses, 
since they all appear in violation of the continuous $\,R$-symmetry (\ref{r}).
Altogether there is here no specific hierarchy problem associated with 
the size of $\,\mu\,$.
In general, irrespective of the supersymmetry breaking mechanism
considered, still unknown
(and generally parametrized using a variety of possible
soft supersymmetry breaking terms),
\,one normally expects the various superpartners not to be too heavy, 
otherwise the corresponding new mass scale introduced in the game 
would tend to contaminate the electroweak scale, thereby
{\it creating\,} a hierarchy problem in the Supersymmetric Standard Model.
Superpartner masses are then normally expected to be naturally of the
order of $\,m_W$, ~or at most in the $\ \sim\,$ TeV$/c^2\,$ range.

\vskip .25truecm

The Supersymmetric Standard Model (``minimal'' or not),
with its $R$-parity symmetry (absolutely conserved, or not),
provided the basis for the experimental searches for the new superpartners
and Higgs bosons, starting with the first searches for gluinos and photinos, 
selectrons and smuons, at the end of the seventies. 
How the supersymmetry should actually be broken,
if indeed it is a symmetry of Nature, is not known yet.
Many good reasons to work on the Supersymmetric Standard Model 
and its various extensions have been discussed, 
dealing with Dark Matter, supergravity, gauge coupling unification,
extended supersymmetry, new spacetime dimensions, 
superstrings, \hbox{``$M$-theory'',} \hbox{... .}
~Despite all the efforts made for more than twenty years 
to discover the new inos and sparticles, 
we are still waiting for experiments to disclose this missing 
half of the SuperWorld.
Still supersymmetry may well be, beyond quantum physics and general relativity,
the next fundamental symmetry to be discovered 
in the physics of fundamental particles and interactions,
enlarging our vision of space and time to the new
anticommuting dimensions of superspace.

\end{document}